\documentclass[12pt,preprint]{aastex}
\usepackage{amsmath, amsthm, amssymb}
\usepackage{natbib}
\usepackage{lscape}
\usepackage{color}

\usepackage{graphicx}

\usepackage{times}

\ifnum 2<1

\fi

\begin{document}

\title{
       Understanding solar cycle variability} 

\author{R.H. Cameron and
        M. Sch\"{u}ssler}
\affil{Max-Planck-Institut f\"ur Sonnensystemforschung\\
               Justus-von-Liebig-Weg 3, 37077 G\"ottingen, Germany}
\email{cameron@mps.mpg.de}

\begin{abstract}
{\bf Version: \today}\\

The level of solar magnetic activity, as exemplified by the number of
sunspots and by energetic events in the corona, varies on a wide range
of time scales. Most prominent is the 11-year solar cycle, which is
significantly modulated on longer time scales. Drawing from dynamo
theory together with empirical results of past solar activity and of
similar phenomena on solar-like stars, we show that the variability of
the solar cycle can be essentially understood in terms of a weakly
nonlinear limit cycle affected by random noise.  In contrast to
ad-hoc `toy models' for the solar cycle, this leads to a generic
normal-form model, whose parameters are all constrained by observations.
The model reproduces the characteristics of the variable solar activity
on time scales between decades and millennia,  including the
occurrence and statistics of extended periods of very low
activity (grand minima).  Comparison with results obtained with a
Babcock-Leighton-type dynamo model confirms the validity of the
normal-mode approach.
\end{abstract}


\section{Introduction}
\label{sec:intro}

\begin{figure}
\begin{center}
\includegraphics[scale=0.8]{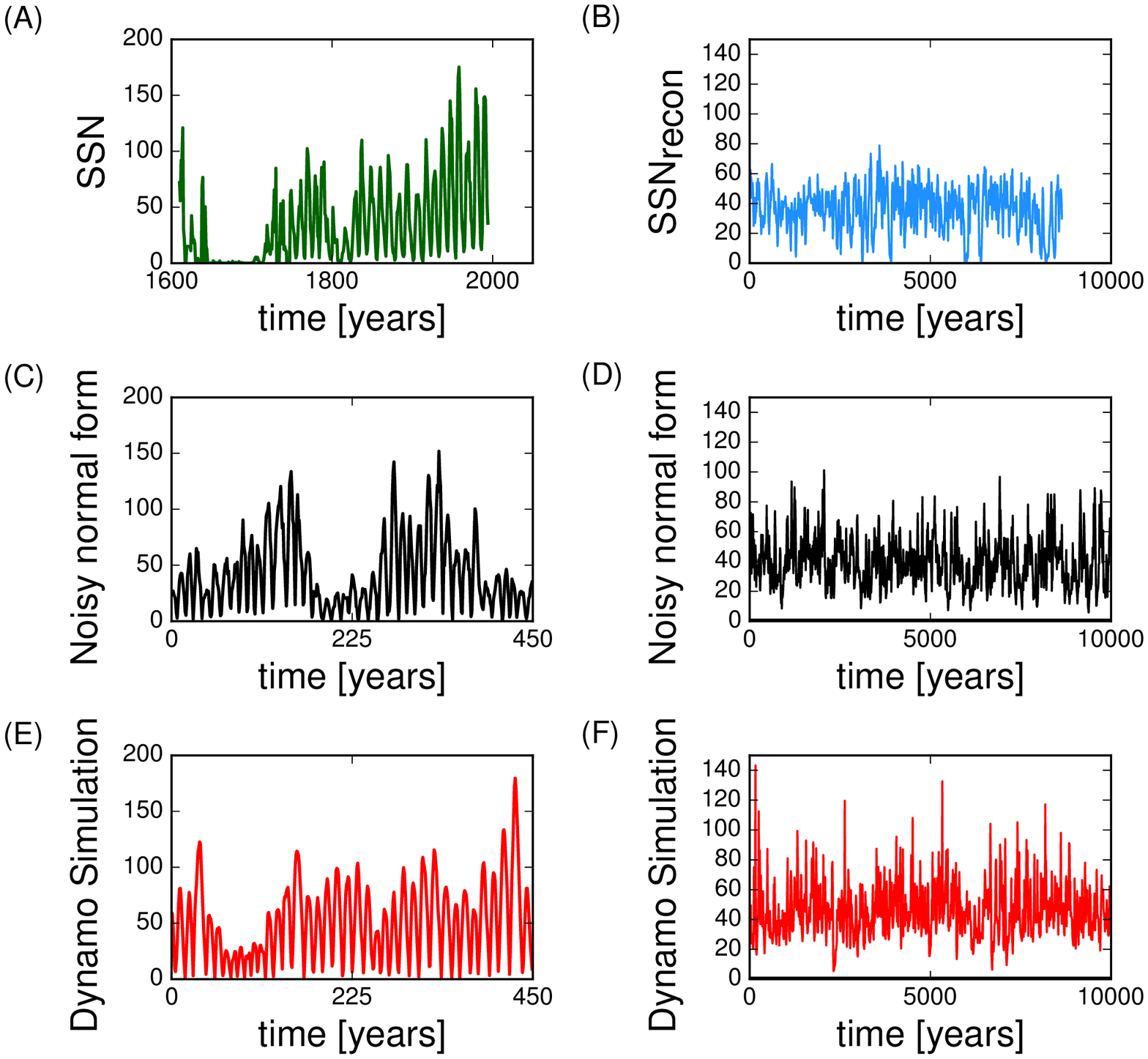}
\caption{Time series of observed and simulated sunspot numbers (SSN). A:
  yearly group sunspot numbers \citep{Hoyt:Schatten:1998} obtained from
  observations between 1610 and 1995. B: sunspot numbers reconstructed
  from cosmogenic isotopes \citep{Usoskin:etal:2016} with 10-year
  resolution back to 9,000 yr BP.  C,D: results for time intervals of
  comparable lengths taken from Monte-Carlo simulations of a weakly
  nonlinear, noisy limit cycle (normal-form model) with parameters
  determined from observations. E,F: results obtained using a
  Babcock-Leighton-type dynamo model with fluctuating sources
  \citep{2017A&A...599A..52C}.}
\label{fig:1}
\end{center}
\end{figure}

 Apart from its 11-year (quasi)periodicity, the most striking
  property of the solar activity record is the marked variability of the
  cycle amplitudes (cf. top panels of Fig.~\ref{fig:1}), including
  extended intervals of very low or particularly high activity
  \citep[grand minima and maxima, see][]{Usoskin:2017}. Understanding
  the nature of the variability is a prerequisite for sensible attempts
  to predict future activity levels.  Therefore, we need to clarify to
  what extent randomness, intrinsic periodicities apart from the 11-year
  cycle, and nonlinearities of the underlying dynamo process generating
  the solar magnetic field contribute to the observed long-term
  variability of solar activity.

  There exists a rich literature describing attempts to understand the
  variability in the framework of hydromagnetic dynamo theory, a full
  review of which is beyond the scope of this paper \citep[see,
  e.g.,][]{Tobias:2002, Charbonneau:2010, Charbonneau:2014}. Such
  studies can be roughly divided into two approaches: (1) nonlinear
  dynamics and deterministic chaos, and (2) random fluctuations of
  dynamo excitation.  The nonlinear dynamics approach typically
  considers the bifurcation structure of low-order dynamical systems
  \citep[see reviews by][]{Weiss:1990, Tobias:etal:1995,
  Lopes:etal:2014}, but models based on nonlinear PDEs have also been
  investigated frequently \citep[e.g.,][]{Schmitt:Schuessler:1989,
  Tobias:1997, Bushby:2006}. Studies assuming random fluctuations reach
  from the minimalistic `model' of \citet{Barnes:etal:1980} to detailed
  considerations of the mode structure of stochastically excited dynamos
  \citep{Hoyng:1993, Hoyng:Vangeffen:1993,
  Ossendrijver:Hoyng:1996}. Most of these studies adopt an
  $\alpha\Omega$-type dynamo approach with random fluctuations of the
  $\alpha$-effect thought to result from the non-stationary nature of
  solar convection \citep[e.g.][]{Choudhuri:1992, Moss:etal:1992,
  Ossendrijver:etal:1996}. There are also studies combining nonlinear
  dynamo models with random fluctuations, either in the framework of
  low-order dynamical systems \citep[e.g.][]{Mininni:etal:2001,
  Passos:Lopes:2011} or assuming detailed dynamo models
  \citep[e.g.][]{Charbonneau:Dikpati:2000, Mininni:Gomez:2002,
  Moss:etal:2008, 2017ApJ...834..133L}. Grand minima can also result from of `on-off
  intermittency' due to the interaction of two spatially separated
  dynamos \citep{Platt:etal:1993, Schmitt:etal:1996, Passos:etal:2014}.
  Recently, \citet{Olemskoy:etal:2013} have estimated the fluctuating
  source term of a Babcock-Leighton-type dynamo considering the observed
  scatter of the tilt angles of sunspot groups. Assuming weakly
  supercritical dynamo excitation, the simulated long-term evolution of
  solar activity exhibits grand minima whose statistics are consistent
  with the actual solar record \citep[see
  also][]{Kitchatinov:Olemskoy:2016}. Global 3D-MHD simulations of
  convection and magnetic field in a rotating spherical shell typically
  show strong variability of dynamo-generated magnetic field and may
  also provide cyclic fields \citep[e.g.][]{Augustson:etal:2015,
  Passos:Charbonneau:2014, Kaepylae:etal:2016, Hotta:etal:2016,
  Fan:Fang:2016}. However such simulations remain far from the
  Sun in terms of various parameters, and so the variability seen in
  the simulations is not yet directly relatable to that of the Sun.

Although no definite quantitative model of the global dynamo exists so
far, the basic ingredients are uncontroversial. The systematic
properties of sunspot groups \citep{Hale:etal:1919} indicate that they
originate from a reservoir of organized East-West orientated (toroidal)
field in the solar convection zone. This field is generated by winding
up a poloidal field (such as a dipole field aligned with the rotation
axis) by the differential rotation of the Sun, so that its axisymmetric
component dominates. The poloidal field is (re)generated against the
effect of Ohmic decay by the collective effect of loops formed from the
toroidal field by convective flows and/or magnetic buoyancy. The loops
become twisted owing to the Coriolis force and thus acquire a systematic
meridional component \citep{Parker:1955, Babcock:1961,
Steenbeck:Krause:1966}. This interplay of toroidal and poloidal magnetic
field leads to a 22-year magnetic cycle and an 11-year cycle of sunspot
activity.

Here we consider these basic ingredients of solar dynamo theory together
with solar and stellar observations to elucidate the nature of the
variability of the solar cycle.  We argue that the dynamo works near
its excitation threshold (marginal state), so that a normal-form model
representing the essence of the underlying dynamo process can be
used. 
This model is generic in the sense that it is valid for any
weakly nonlinear system in the vicinity of a supercritical Hopf
bifurcation, independent of the nature of the nonlinearity. The very
setup of the model and its parameters are all constrained by
observations.

\section{Generic normal-form model}
\label{sec:normal_form}
 
Irrespective of the details of most models for the solar dynamo proposed
so far, the corresponding systems of equations for the magnetic field
components show qualitatively similar behaviour near the onset of dynamo
action, which is governed by a control parameter (often called `dynamo
number') involving differential rotation, Coriolis effect, and magnetic
diffusion. As long as the control parameter remains below a critical
value, the stationary solution has zero magnetic field and the dynamo is
not excited. When the critical value of the control parameter is
exceeded, the system exhibits a periodic solution. This behaviour is
generic for almost all models of the global solar dynamo (technically
named $\alpha\Omega$ dynamos). In the language of dynamical systems, the
periodic solution emerges from a (supercritical) Hopf bifurcation: a
fixed point becomes unstable and spawns a limit cycle
\citep{Guckenheimer:Holmes:1983, Tobias:etal:1995}.

The study of solar-like stars has revealed that the level of magnetic
activity systematically declines with decreasing stellar rotation rate
\citep[e.g.,][]{Reiners:2012,Reiners:etal:2014}. In particular, the
relatively slow rotation rate of the Sun appears to put it near to the
threshold for which global dynamo action ceases
\citep{vanSaders:etal:2016, Metcalfe:etal:2016}. This allows us,
irrespective of the nature of the nonlinearity that limits the amplitude
of the cycle, to describe the solar dynamo generically by the normal
form of a weakly nonlinear system near a Hopf bifurcation,  which is
independent of the nature of the nonlinearity \citep{Arnold:1972,
Guckenheimer:Holmes:1983}, viz.
\begin{eqnarray}
  \frac{\mathrm{d} X}{\mathrm{d}t}-(\beta+i\omega_0) X +(\gamma_r+i
  \gamma_i) |X|^2 X =0 \,,
  \label{eqn:1}
\end{eqnarray}
where $X$ is a complex variable. Its real and imaginary components are
related to the toroidal and poloidal components of the magnetic field in
a manner that depends on the kind of nonlinearity considered (e.g., back
reaction on the differential rotation or quenching of the regeneration
term for the poloidal field). We may therefore consider the real or the
imaginary part of $X$ to represent the cyclically varying field
magnitude. For simplicity, we take the activity level as quantified by
the sunspot number, SSN, to be proportional to the absolute value of the
field and write ${\rm SSN} = |{\rm Re}(X)|$, thus scaling $|{\rm
Re}(X)|$ in units of the sunspot number. This quantity represents the
11-year sunspot cycle, in contrast to the 22-year magnetic cycle
reflected in ${\rm Re}(X)$ and ${\rm Im}(X)$.

The quantity $\beta$ in Eq.~(\ref{eqn:1}) is the linear growth rate of
the cycle amplitude (related to the dynamo excitation, i.e.,
supercriticality of the dynamo number) and $\omega_0$ is the (magnetic)
cycle frequency in the limit of zero amplitude. The parameters
$\gamma_r$ and $\gamma_i$ generically represent the nonlinearity of the
system. They determine the cycle amplitude, $|X|=\sqrt{\beta/\gamma_r}$,
and the nonlinear cycle frequency, $\omega = \omega_0 - \gamma_i
\beta/\gamma_r$.

In order to study the variability of the cycle amplitude, we need to
account for the randomness inherent to the dynamo process.  There is
strong evidence that the dynamo is of Babcock-Leighton type
\citep{Wang:Sheeley:2009,Munoz:etal:2013}. This means that magnetic flux
connected to the polar fields (axial dipole) is the relevant poloidal
field for the generation of the toroidal field by differential rotation
in the convection zone \citep{Cameron:Schuessler:2015}. The polar fields
result from the emergence of bipolar magnetic regions (sunspot groups)
with a systematic average tilt relative to the solar East-West direction
together with the subsequent transport of their magnetic flux across the
surface by differential rotation, supergranulation, and meridional flow
\citep{Mackay:Yeates:2012, Wang:2016}. Consequently, any scatter in the
properties of the bipolar regions (e.g., emergence latitude, tilt angle)
or in the flux transport process introduces a corresponding scatter in
the resulting poloidal dipole field and, therefore, in the amplitude the
subsequent cycle \citep{Charbonneau:Dikpati:2000, Wang:Sheeley:2009,
Jiang:etal:2014}. Since Babcock-Leighton dynamos can be put into the
general mathematical framework of $\alpha\Omega$-dynamos
\citep{Stix:1974}, we can describe this effect by random scatter in the
term generating the poloidal field, which enters the normal form model
in the form of multiplicative noise. Eq.~(\ref{eqn:1}) thus transforms
into a stochastic differential equation, viz.
\begin{eqnarray}
  \mathrm{d} X = \left(\beta+i\omega_0 -(\gamma_r + i \gamma_i) |X|^2
  \right)X \mathrm{d}t +\sigma X \mathrm{d}W_c =0 \,,
\label{eqn:2}
\end{eqnarray}
where we take $W_c$ to represent a complex Wiener process (random walk
with uncorrelated, Gaussian distributed increments) with a variance of
unity after 11 years.  The real parameter $\sigma$ then corresponds to
the standard deviation of the cycle amplitudes due to the noise in the
generation process.

Values for all five parameters entering Eq.~(\ref{eqn:2}) are
constrained by empirical results. The linear growth rate, $\beta$, can
be estimated from the time scale for the recovery from a grand minimum
of very low activity. The duration of the Maunder minimum of about 70
years implies a recovery time of the order of decades, so that we take
$\beta=1/50$~year$^{-1}$ as a reference value. The activity cycle
appears to have persisted at low amplitude and with unchanged period
during the Maunder minimum \citep{Beer:etal:1998, Vaquero:etal:2015}, so
that we take the (magnetic) cycle frequency to be unaffected by the weak
nonlinearity, $\omega=\omega_0= 2\pi/22\,$year$^{-1}$, and thus
$\gamma_i=0$. In order to fix the parameter $\gamma_r$, we need to
determine the (average) cycle amplitude, $\sqrt{\beta/\gamma_r}$, in
terms of the sunspot number. For a cycle of sinusoidal shape, the
amplitude is given by $\pi/2$ times its mean level. Using the average of
the Group Sunspot Numbers \citep{Hoyt:Schatten:1998} since 1700, the end
of the Maunder minimum, we obtain a cycle amplitude of 64 and thus
$\gamma_r=4.9\cdot10^{-6}$~year$^{-1}$.

The remaining parameter to be fixed is the noise level, $\sigma$, 
 which we relate to the scatter of the tilt angles of bipolar magnetic
 regions. Surface flux transport simulations show that the observed
 Gaussian scatter of the tilt angles leads to a fluctuation level of
 30\%--40\% of the polar dipole field
 \citep{Jiang:etal:2014,Munoz:etal:2013}. This implies a degree of
 fluctuations of the cycle amplitudes owing to the linear correlation
 between the polar field at the end of a cycle and the amplitude of the
 next cycle \citep{Wang:Sheeley:2009}. This level is consistent with the
 value of about 35\% for the scatter of the cycle amplitudes determined
 from the sunspot numbers observed since 1700 amounts and also with the
 scatter shown by the 10-year sampled sunspot numbers reconstructed from
 the cosmogenic isotopes \citep{Usoskin:etal:2016}.
Since the limit cycle is an attractor, we have a slight damping of the
random perturbations during a cycle, so that we have to use a somewhat
higher value, $\sigma=0.40$, in order to compensate for this effect and
reproduce the observed cycle-to-cycle fluctuation of 35\%.  The effect
on the results of varying the parameters $\beta$ and $\sigma$ is
discussed further below.

\begin{figure}
\begin{center}
\includegraphics[scale=0.6]{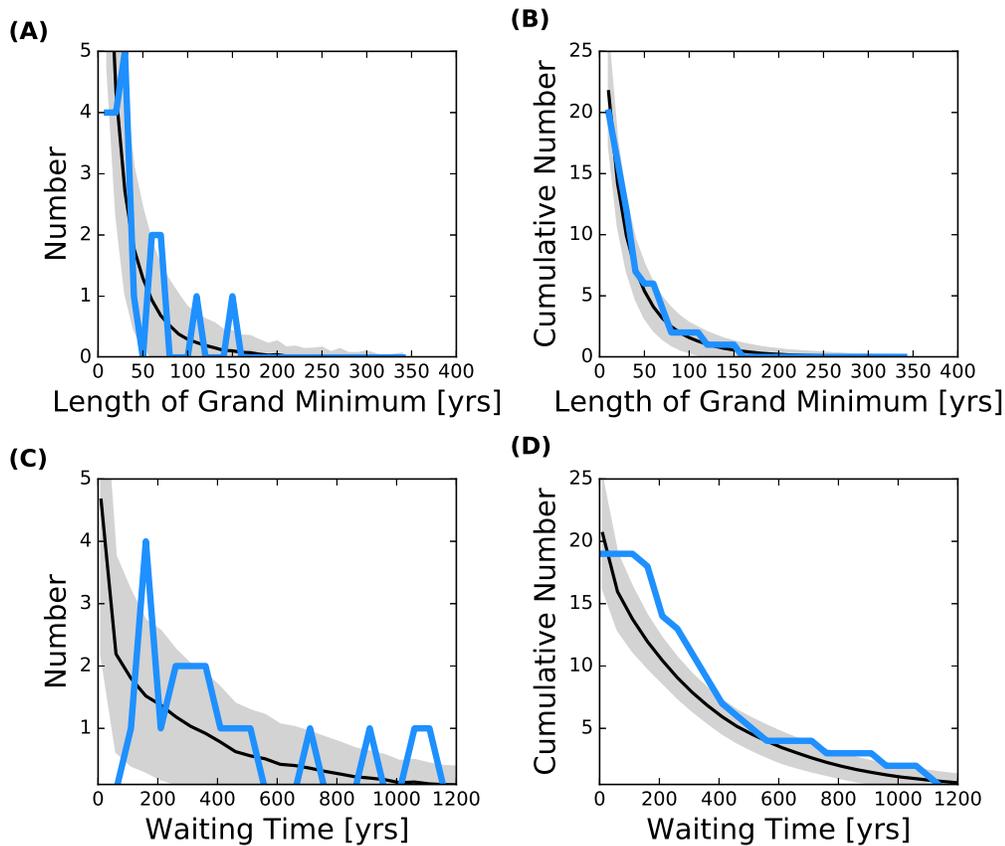}
\caption{Statistics of grand minima, extended periods of very low solar
  activity (here defined as sunspot number below 20 for at least two
  consecutive 10-year averages). A: Distribution of the lengths of grand
  minima. B: Corresponding cumulative distribution. C: Distribution of
  the waiting times between grand minima. D: Corresponding cumulative
  distribution. Empirical distributions derived from the cosmogenic
  isotope record \citep{Usoskin:etal:2016} are given by the blue lines
  while the results from the normal-form model (based on 1000
  realizations of 10,000 years length each) are shown in black with grey
  areas indicating the standard deviation. The numbers given refer to an
  interval of 8410 years as covered by the reconstruction from the
  cosmogenic isotope record.}
\label{fig:2}
\end{center}
\end{figure}

We used the Euler-Maruyama method \citep{Kloeden:Platen:1992} to perform
Monte-Carlo simulations of Eq.~(\ref{eqn:2}) with the parameter values
given above and a time step of one day. Random numbers were generated
with the routine `random.normal' from the Python package `numpy' version
1.10.4.  The convergence of the numerical result was checked against the
analytic solution of the normal form without noise.  Covering 10,000
years for each realization, we obtained the amplitude variability over a
large range of time scales.  Results for one such realization are
illustrated in the middle panels of Fig.~\ref{fig:1}. Panel C covers a
period of 450 years, exhibiting extended periods of very low activity
(grand minima). Panel D shows the 10-year averages for the whole
time interval of 10,000 years. The main features of the actual and the
simulated time series are qualitatively similar: there is considerable
variability of the cycle amplitudes with occasional grand minima (and
maxima).  
For a more quantitative comparison, Fig.~\ref{fig:2} gives the
  non-cumulative (left panels) and cumulative (right panels)
  distributions of the lengths of grand minima and of the waiting times
  between the end of one grand minimum and the start of the next from
  the simulations (black lines) in comparison to the empirical
  distributions (blue lines) derived from the cosmogenic isotope record
  \citep{Usoskin:etal:2016}. The model distributions are well
  approximated by exponentials, which would be expected for a Poisson
  process \citep[see also][]{Olemskoy:etal:2013}. Overall, the
  distributions appear to be consistent with each other. Whether the
  empirical events in the tail of the waiting time distribution
  represent a significant deviation from an exponential distribution
  \citep[cf.][]{Moss:etal:2008} cannot be decided owing to the small
  number (20) of grand minima in the isotope record. Note that the
  normal-form model generically shows ongoing low-amplitude cycles
  during grand minima, such as indicated during the Maunder minimum
  \citep{Beer:etal:1998, Vaquero:etal:2015}. 

\begin{figure}
\begin{center}
\includegraphics[scale=0.6]{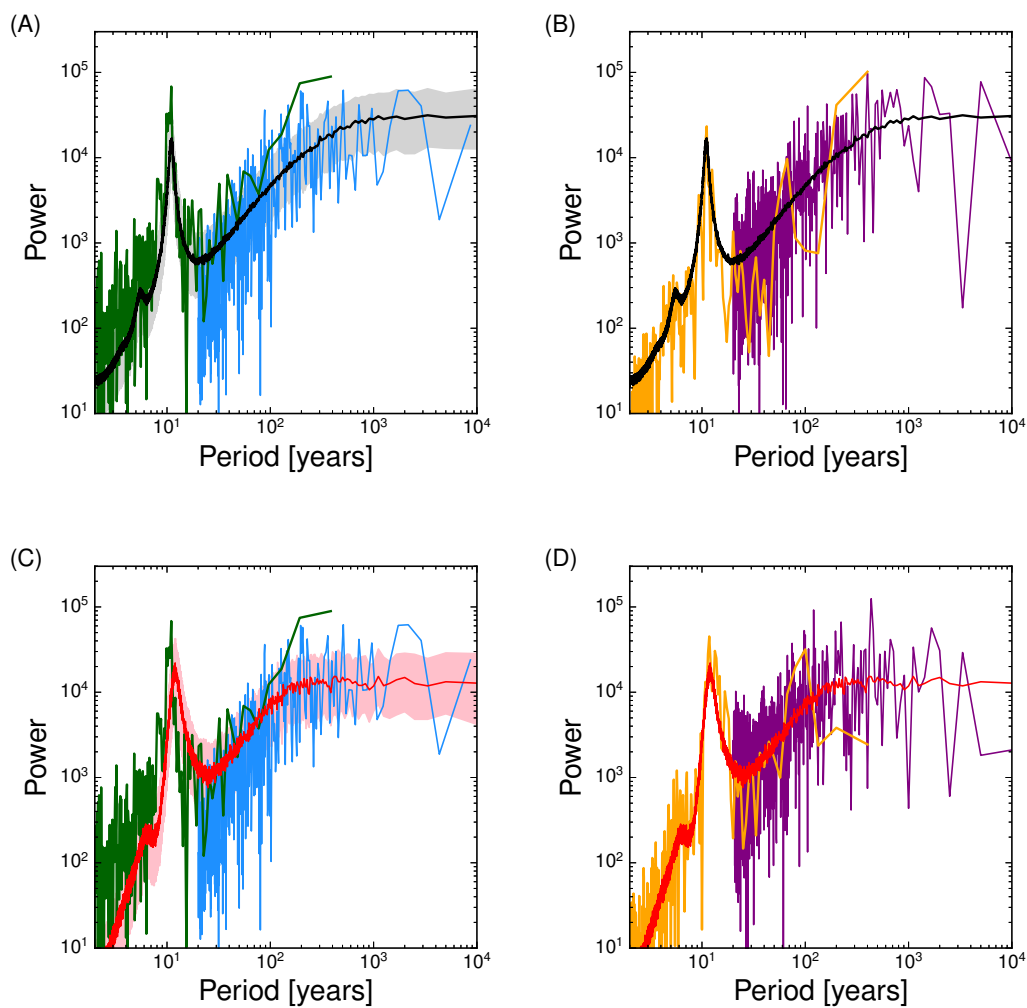}
\caption{Power spectra of observed and simulated sunspot numbers.  A:
  spectra corresponding to the empirical data displayed in Panels A and
  B of Fig.~\ref{fig:1} (green line: yearly sunspot numbers; blue
  line: reconstructed sunspot numbers from cosmogenic isotopes) together
  with the median (black line) and the range covering the 25\% and 75\%
  quartiles (grey band) for 1000 simulations of the noisy limit
  cycle. B: corresponding spectra for a single realization of the
  model. C: empirical spectra in comparison to those obtained with
  Babcock-Leighton-type dynamo model with fluctuating sources (red line:
  median of 120 realizations; pink band: range between the 25\% and 75\%
  quartiles). D: spectra for one realization of the Babcock-Leighton
  dynamo model.}
\label{fig:3}
\end{center}
\end{figure}

Further quantitative comparison between model and data is obtained by
considering power spectra as shown in Fig.~\ref{fig:3}. Panel A
corresponds to the historical sunspot record (yearly resolution, green
line) and to the reconstruction based on cosmogenic isotopes (10-year
resolution, blue line) in comparison to the median of 1000~realizations
of 10,000~years each of the normal-form model (black line). The model
results are consistent with the observed spectra, demonstrating that a
weakly nonlinear limit cycle driven by random noise is sufficient to
explain the observed variability of solar activity on a wide range of
time scales.  A single realization of the model is considered in
panel B of Fig.~\ref{fig:3}, which shows spectra for 450 years with
yearly resolution (orange line) and for 10,000 years with 10-year
resolution (purple line). These spectra are qualitatively similar to the
observed spectra shown in panel A, including some apparent (but
spurious) long-term periodicities.

We also considered the effect of varying the parameters $\beta$ (linear
growth rate, related to the supercriticality of the dynamo) and
$\sigma$, the level of the random forcing. They represent two competing
effects: the forcing tends to drive the solution away from the limit
cycle while perturbations of the stable limit cycle decay with a
timescale $(2\beta)^{-1}$. Changing these parameters therefore varies
the relative importance of the two effects.  In addition to the
reference value, $\beta=1/50$~year$^{-1}$, suggested by the recovery of
the Sun from the Maunder minimum, we considered values representing a
much smaller growth rate, $\beta=1/250$~year$^{-1}$, and a markedly
higher growth rate, $\beta=1/10$~year$^{-1}$. Likewise, we took values
for the random forcing of $\sigma = 0.$ (no perturbations) and $\sigma =
1.$ (very strong perturbations), in addition to the reference value of
$\sigma = 0.4$ indicated by the observed variability of the solar cycle
maxima.

\begin{figure}
\begin{center}
\includegraphics[scale=0.8]{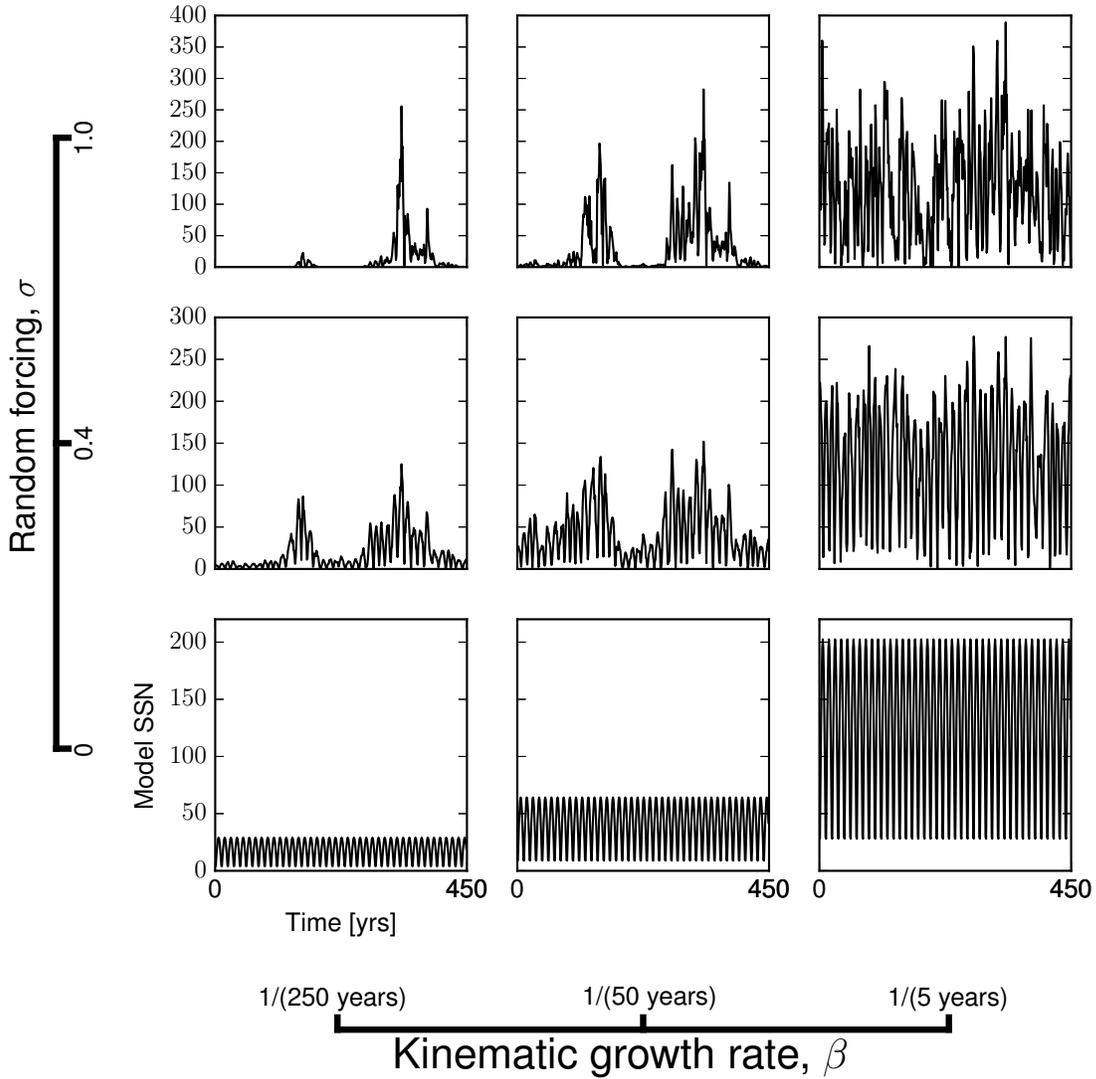}
\caption{Normal-form model: effect of varying the strength of the random
  forcing, $\sigma$, and the kinematic growth rate, $\beta$. Shown are
  activity levels (scaled in terms of sunspot numbers) for one
  realization each, covering arbitrarily chosen time intervals of
  450~years (roughly the length of the empirical sunspot number
  record). The central panel refers to the `solar' reference case. }
\label{fig:4}
\end{center}
\end{figure}

\begin{figure}
\begin{center}
\includegraphics[scale=0.8]{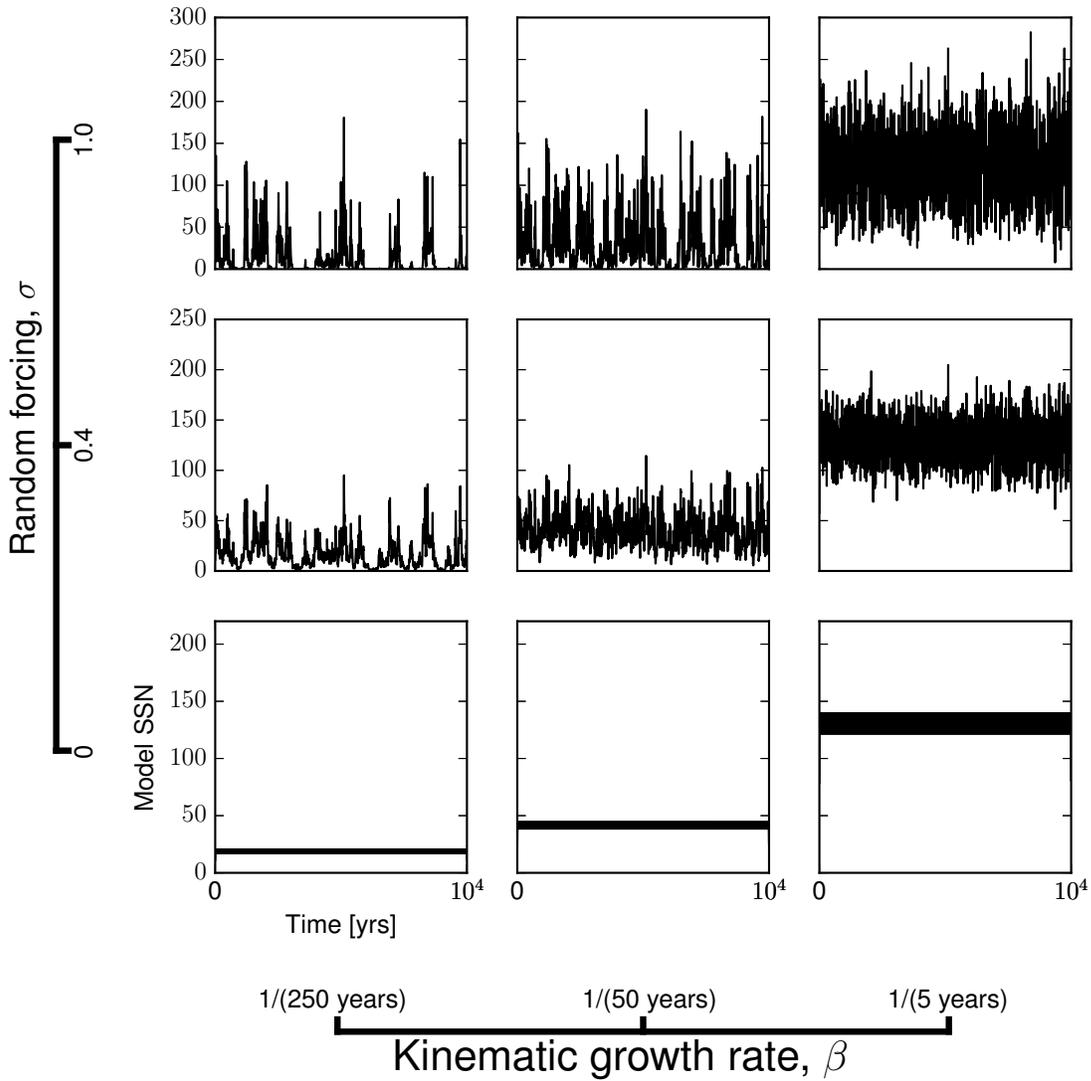}
\caption{Similar to Fig.~\ref{fig:4}, but for 10-year
  averages covering the full length of 10,000 years of the same
  realizations.  The time covered roughly corresponds to the length of
  the reconstruction of solar activity from cosmogenic isotope data.}
\label{fig:5}
\end{center}
\end{figure}

For the cases with $\beta=1/50$~year$^{-1}$ and $\beta=1/10$~year$^{-1}$
we performed 1000 simulations of 11,000~years length each. We excluded
the first 1,000 years from each realization in order to stay clear of
the transients related to the arbitrary initial conditions. For the very
weakly excited cases with $\beta=1/250$~year$^{-1}$, we run 1000
simulations of 21,000~years length each, excluding the first 11,000
years from the analysis.  To illustrate the results, a stretch of 450
years of one arbitrarily chosen realization for each pair of parameters
is shown in Fig.~\ref{fig:4}, while 10-year running averages covering
10,000 years are given in in Fig.~\ref{fig:5}. In the cases without
random forcing (bottom rows of the figures), the solutions are rectified
sine functions with amplitude $\sqrt{\beta/\gamma_r}$.  For the
observationally well constrained reference value of the random forcing,
$\sigma=0.4$, the cycles show variations in amplitude and phase (middle
rows in Figs.~\ref{fig:4} and \ref{fig:5}). The value of $\beta$
determines the rate of occurrence and the lengths of grand minima: since
the recovery time from perturbations is $(2\beta)^{-1}$, low values of
$\beta$ lead to long extended minima, while high values suppress their
occurrence. In the case of very strong random forcing (top rows of
Figs.~\ref{fig:4} and \ref{fig:5}), the perturbations lead to high
cycle-to-cycle fluctuations and to very long grand minima for low
$\beta$.  From these results, we would expect that very slowly rotating
stars, which are even nearer to critical dynamo excitation, show very
long quiescent phases, while stars with higher excitation (faster
rotators) display very strong cycle-to-cycle fluctuations and, if at
all, only rarely show short grand minima of activity.

Fig.~\ref{fig:3} shows median power spectra corresponding to 1000
realizations for each of the nine parameter combinations. The case
without perturbations (bottom row) obviously shows the spectra of a
rectified sine wave. For growing strength of the random forcing, the
11-year peak becomes less prominent and broader as the amplitude and
phase fluctuations of the cycles grow stronger.  For growing dynamo
excitation (higher $\beta$), the long-period tail of the spectrum
flattens as a result of the shortening and increasing suppression of the
grand minima.

\section{Babcock-Leighton-type dynamo model}
\label{sec:BL}

To demonstrate that the normal-form approach actually fits in the
context of a more detailed nonlinear dynamo model, we consider an
updated version of the Babcock-Leighton dynamo
\citep{Babcock:1961,Leighton:1969,Wang:etal:1991}, which reproduces key
features of the solar cycle \citep{2017A&A...599A..52C}.  The model
comprises the essential ingredients of a flux-transport dynamo
\citep{Charbonneau:2010, Charbonneau:2014}. It considers the
axisymmetric part of the magnetic field and is based on the evolution
equations for the azimuthal component of the vector potential
(determining the poloidal field) at the solar surface,
\begin{equation}
a(\theta,t)=\frac{1}{\sin \theta}
\int_0^{\theta} \sin \theta {R^2_{\odot}} B_r|_{R=R_\odot}
{\mathrm{d}}\theta \,,
\end{equation}
and the radially integrated toroidal flux per
radian, 
\begin{equation}
b(\theta,t)=\int_{R_b}^{R_\odot} B_\phi r \mathrm{d}r \,,
\end{equation}
both as functions of colatitude, $\theta$, and time.  Here, $R_\odot$ is
the solar radius, $R_b$ the radial location of the bottom the solar
convection zone. $B_r$ and $B_\phi$ are the radial and azimuthal components,
respectively, of the magnetic field. Introducing fluctuations in the
source term for the poloidal field, the evolution equations given in
\citet{2017A&A...599A..52C} become stochastic differential equations,
viz.
\begin{eqnarray}
  \mathrm{d}a&=&
  -\frac{U(\theta)}{R_\odot\sin\theta} 
   \frac{\partial (a \sin\theta)}{\partial \theta} \mathrm{d}t
   +\frac{\eta_{R_\odot}}{R_\odot^2} \frac{\partial}{\partial \theta}
  \left[\frac{1}{\sin\theta}\frac{\partial (a\sin\theta)}{\partial\theta}  
  \right] \mathrm{d}t \nonumber  \\ \noalign{\vspace{0.2cm}}
   & & + \; a_{\rm S}(\theta,t)\mathrm{d}t 
  +\sigma^*  a_{\rm S}(\theta,t) \mathrm{d}W(t,\theta)\,,
\label{eqn:a}
\end{eqnarray}
and 
\begin{eqnarray}
  \label{eqn:b2}
  \mathrm{d}b
  &=& \frac{\partial a \sin\theta}{\partial \theta}
    \left( \Omega_{R_\odot}- \Omega_{R_{\rm NSSL}} \right) \mathrm{d}t
    -\left(\frac{\partial \Omega_{R_{\rm NSSL}}}{\partial \theta}\right) 
    a \sin\theta \mathrm{d}t \nonumber \\ \noalign{\vspace{0.2cm}}
  & & - \; \frac{1}{R_\odot} 
    \frac{\partial(V_0\sin(2\theta)b)}{\partial\theta} \mathrm{d}t
    + \frac{\eta_0}{R_\odot^2}  \frac{\partial}{\partial \theta} 
    \left[ \frac{1}{\sin \theta} \frac{\partial}{\partial \theta} 
    \left(b\sin\theta \right) \right] \mathrm{d}t \,.
\end{eqnarray}
Here $\Omega_{R_\odot}(\theta)$ and $\Omega_{R_{\rm NSSL}}(\theta)$ are,
respectively, the angular rotation rates at the solar surface and at the
base of the near-surface shear layer, for which we take the same
profiles as in
\citet{2017A&A...599A..52C}. $\eta_{R_{\odot}}=350$~km$^2
\cdot$s$^{-1}$ and $\eta_0$ are, respectively, the magnetic
diffusivities at the solar surface and in the bulk of the convection
zone.  {\rm $U(\theta)$ represents the poleward meridional flow at the
surface, for which we use the profile given by \citet[][see their
Eqs.~9--11]{Hathaway:rightmire:2011}. $V_0 \sin(2 \theta)$ refers to the
equatorward return flow affecting the toroidal field in the bulk of the
convection zone; we take $V_0=2$~m$\cdot$s$^{-1}$, which roughly
corresponds to the speed of equatorward propagation of the activity
belts.}  $a_{\rm S}$ is the source of the poloidal field resulting from
the emergence of bipolar magnetic regions, which we write as
\begin{equation}
  a_{\rm S}(\theta,t) = \frac{\alpha_0}{1+b^2/b_c^2} \cos\theta\sin\theta 
  \, b(\theta,t)\,,
\label{eq:alpha}
\end{equation}
where we have introduced a nonlinearity with parameter $b_c$.
The quantities $\alpha_0$,
$\eta_0$, $b_c$ and $\sigma^*$ are the free parameters of the model.

The random forcing is considered as a two-dimensional Wiener process,
$W(t,\theta)$, which depends on both latitude and time and has a
variance of 1 radian$^{-1}$ after 11 years. The strength of the forcing
is determined by the parameter $\sigma^*$. Since the noise
parameter, $\sigma$, in the normal-form model is independent of the
growth rate (dynamo excitation), for consistency we take
$\sigma^*\alpha_0={\rm const.}$ to represent a fixed noise level.  In
the computations, we used a second-order centered difference scheme for
the spatial derivatives with 180 grid points in colatitude, and advanced
the solution in time using the Euler-Maruyama method with a time step of
1 day.  The numerical results for mildly supercritical, nonlinear dynamo
action with this model ($\eta_0=65$~km$^2 \cdot$s$^{-1}$,
$\sigma^*\alpha_0=0.17$~m$\cdot$s$^{-1}$, $b_c=10^{24}$~Mx, and
$\alpha_0 =5.7 \alpha_{\mathrm{crit}}=2.5$~m$\cdot$s$^{-1}$ 
where $\alpha_{\mathrm{crit}}$ is the critical value
of $\alpha$ for the onset of dynamo action with the other parameters as stated),
are shown in the bottom panels of Figs.~\ref{fig:1} and \ref{fig:3}.
The measure of the activity shown here is the integrated subsurface
 toroidal field corresponding to the dipole mode of the dynamo
 $\vert \int_{0}^{90} b \mathrm{d}\theta -\int_{90}^{180} b\mathrm{d}\theta \vert$,
 scaled to a similar level as the sunspot number.  
They are consistent
with the results from observations and from the generic noisy normal-form model.

We also carried out a parameter study with this dynamo model.
There is no simple relationship between the four free parameters of the
Babcock-Leighton model ($\alpha_0$, $\eta_0$, $b_c$ and $\sigma^*$) and the
four parameters of the normal-form model ($\omega_0$, $\beta$, $\gamma_r$ and
$\sigma$). We have chosen parameters so that
the solutions have a period of about 22 years (i.e. an 11-year activity cycle)
and kinematic growth rates and levels of noise similar to those of the
cases presented in the previous section for the normal-form model.
This specifies three of the four degrees of freedom. The fourth constraint
is equivalent to specifying the amplitude of the limit cycle. In this illustrative
study we have kept $b_c=10^{24}$~Mx and expect solutions of this order of magnitude --
the exact amplitude of the limit cycle will however also depend on the other parameters
in a non-trivial way.

\begin{figure}
\begin{center}
\includegraphics[scale=0.8]{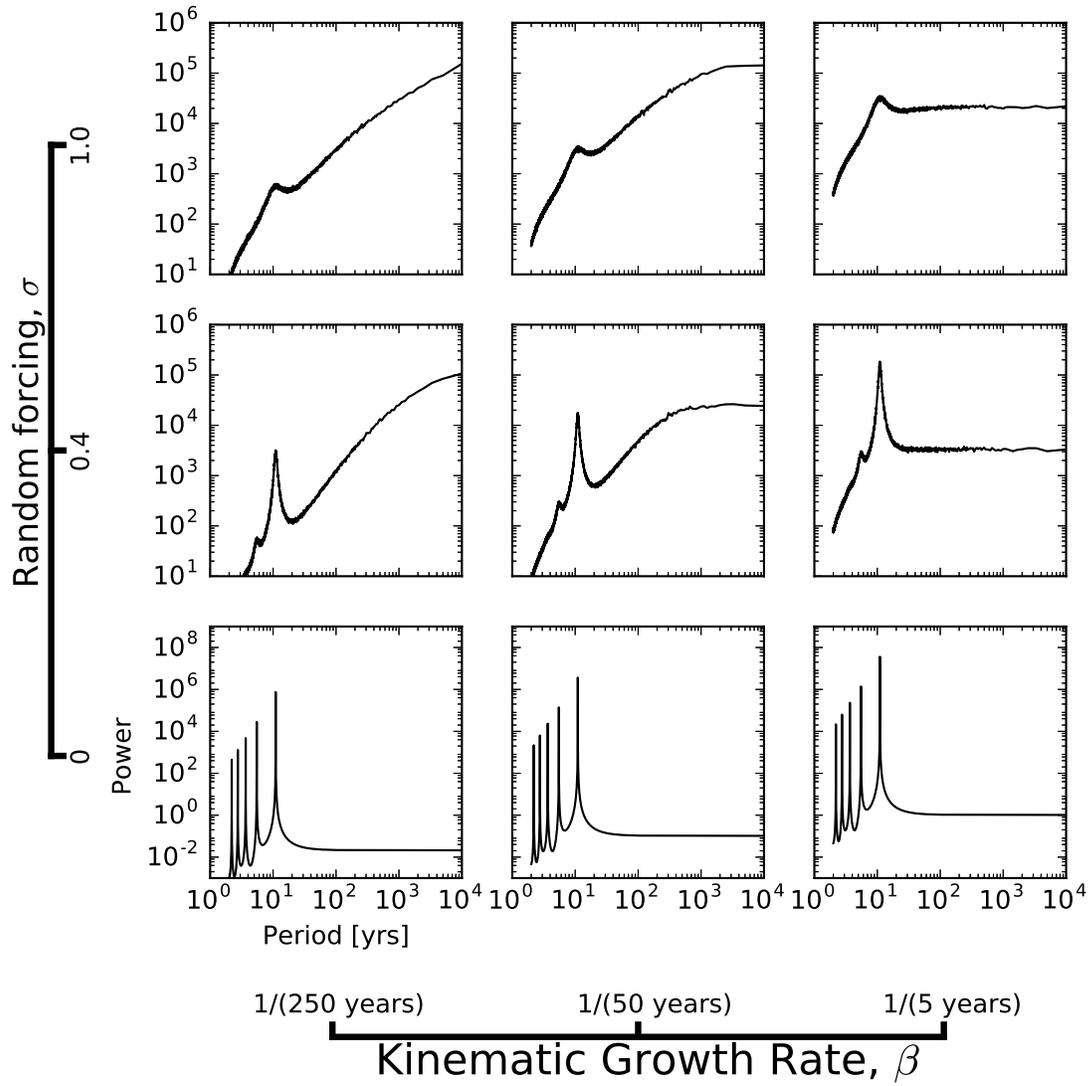}
\caption{Normal-form model: median power spectra for 1,000 realizations
  of 10,000 years length each for the nine combinations of the
  parameters $\sigma$ and $\beta$ considered in
  Figs.~\ref{fig:4} and \ref{fig:5}.}
\label{fig:6}
\end{center}
\end{figure}

\begin{figure}
\begin{center}
\includegraphics[scale=0.8]{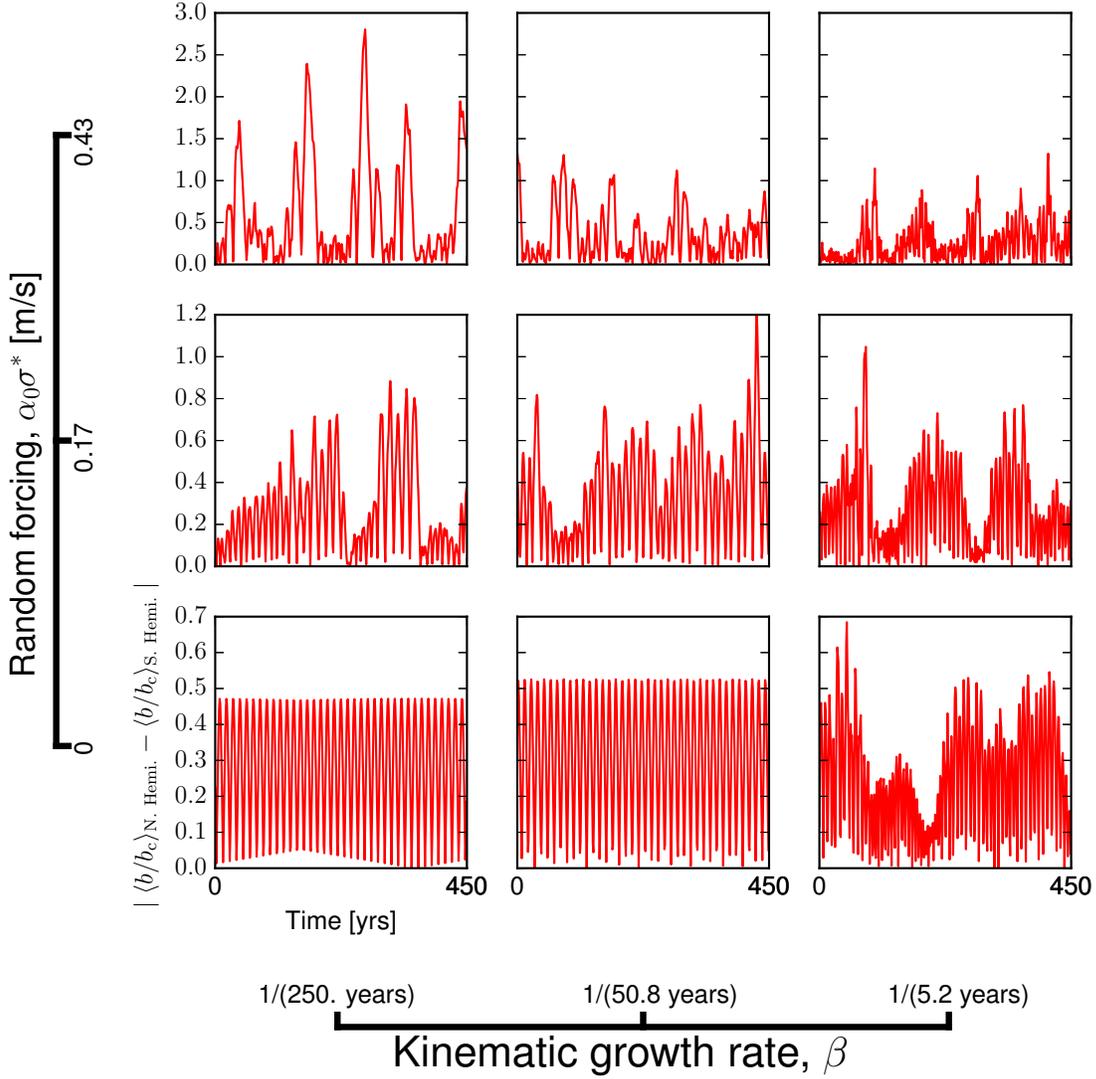}
\caption{Similar to Fig.~\ref{fig:4}, but for the nonlinear
  Babcock-Leighton-type dynamo model. The (effective) strengths of the
  random forcing and the kinematic growth rate are comparable to the
  corresponding cases considered for the normal-form model. Note that
  instead of rescaling to sunspot numbers, the field amplitude is given
  here in terms of the difference between the signed latitudinal average of
  $b/b_c$ in each hemisphere (which reflects the strength of the
  dipole mode of the dynamo).}
\label{fig:7}
\end{center}
\end{figure}

\begin{figure}
\begin{center}
\includegraphics[scale=0.8]{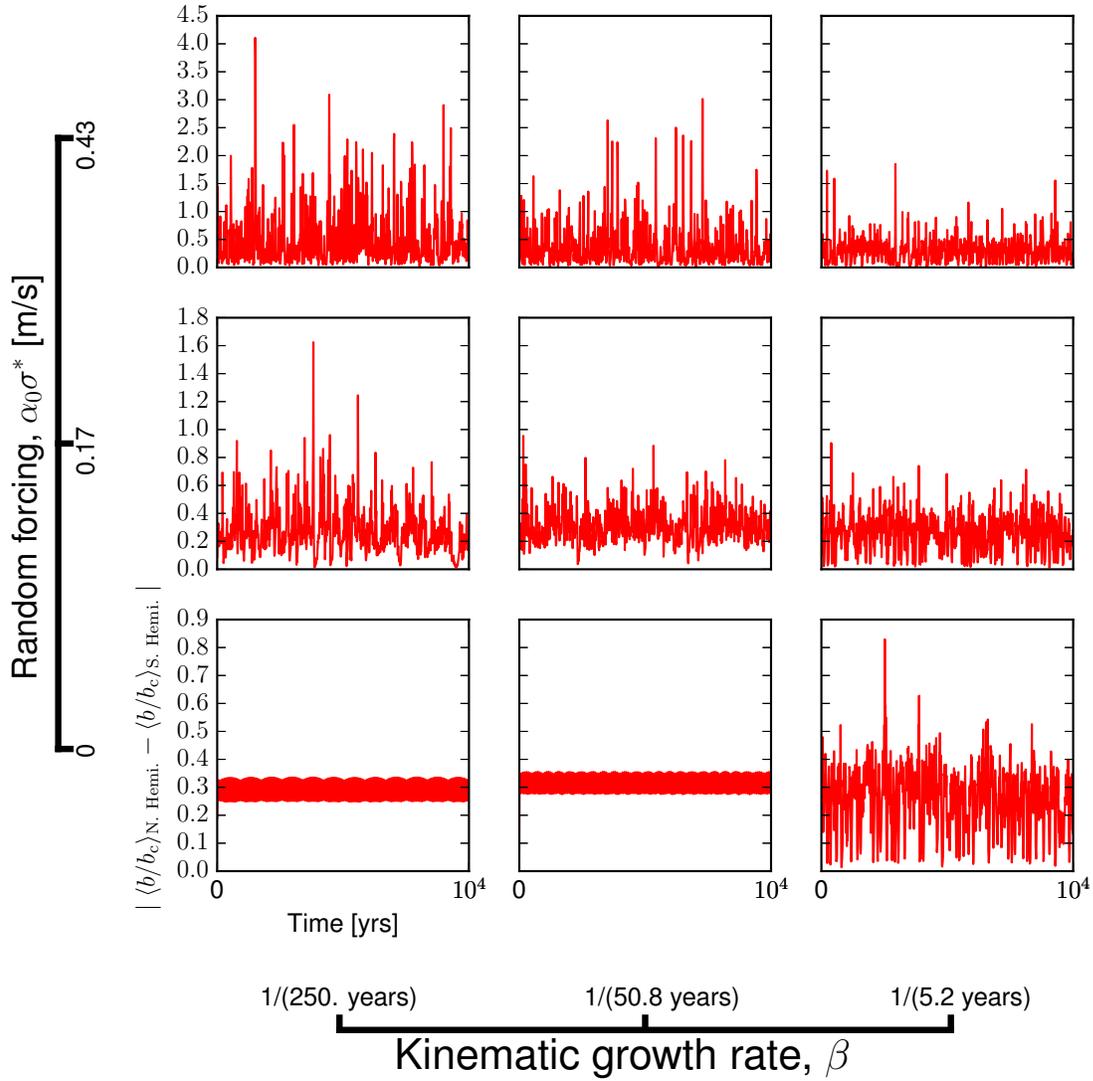}
\caption{Similar to Fig.~\ref{fig:7}, but for 10-year averages
  covering the full length of 10,000 years of the same realizations. The
  corresponding results for the normal-form model are shown in
  Fig.~\ref{fig:5}.}
\label{fig:8}
\end{center}
\end{figure}

\begin{figure}
\begin{center}
\includegraphics[scale=0.8]{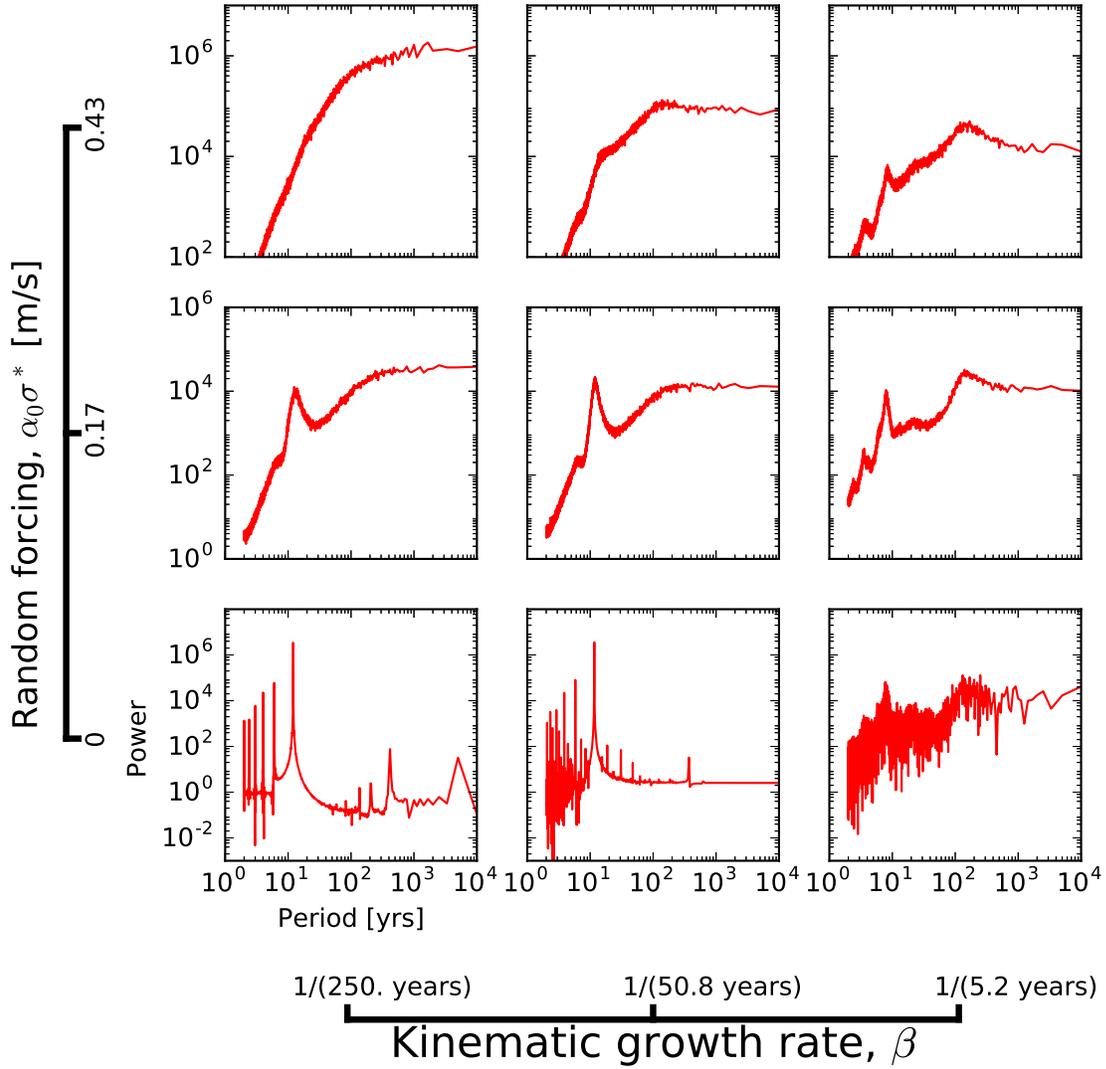}
\caption{Median power spectra for 150 realizations of 10,000 years
  length each of the nonlinear Babcock-Leighton dynamo model. The
  corresponding results for the normal-form model are shown in
  Fig.~\ref{fig:6}.}
\label{fig:9}
\end{center}
\end{figure}

The results of the parameter study are illustrated in
Figs.\ref{fig:7}--\ref{fig:9}, which show the corresponding quantities
in the same format as in Figs.~\ref{fig:4}--\ref{fig:6} for the
normal-form model.  For cases with a very low kinematic growth rate of
1/250~year$^{-1}$ (left column of the figures) we took
$\alpha_0 =1.8 \alpha_{\mathrm{crit}}=1.66$~m$\cdot$s$^{-1}$ and
$\eta_0=75$~km$^2\cdot$s$^{-1}$. For the cases with a growth rate of
1/50.8~year$^{-1}$ (middle column) we chose
$\alpha_0 \bf =5.7 \alpha_{\mathrm{crit}}=2.5$~m$\cdot$s$^{-1}$ and
$\eta_0=65$~km$^2\cdot$s$^{-1}$. Finally, for the case with a high
growth rate of 1/5.2~year$^{-1}$ (right column) we used
$\alpha_0=167 \alpha_{\mathrm{crit}}=20$~m$\cdot$s$^{-1}$ and $\eta_0=30$~km$^2\cdot$s$^{-1}$.  For
the purpose of a qualitative comparison, it is unnecessary to perform a
tedious fine tuning of the parameters in order to exactly match the
growth rates to those of the normal-form model. The cases without random
forcing are shown in the bottom row of
Figs.~\ref{fig:7}--\ref{fig:9}. The forcing for the cases given in the
middle row reproduces the observed variability of the sunspot maxima
since 1700 for the reference case (central panel of the figures). For
the cases with very strong forcing (top row of the figures) we
multiplied the reference value by the same factor 2.5 as in the case of
the normal-form model. In all cases, we carried out 150 simulations covering a time of
11,000~years, omitting the first 1,000~years from the
analysis in order to exclude initial transients.

Comparing the results for both models given in
Figs.~\ref{fig:4}--\ref{fig:6} (normal-form model) and
Figs.~\ref{fig:7}--\ref{fig:9} (dynamo model), respectively, we find
that they are similar for the cases with low and medium growth rates,
including the "solar" reference models (central panels of the
figures).
Because we have varied the two of the model parameters for the
dynamo cases with growth rates of 1/250~years, 1/50.8~years and 1/5.2~years
the amplitudes changes between the different is difficult to interpret.
While the individual realizations obviously differ in detail
between the normal form and dynamo models,
the average spectra are very similar. For the high growth rate, however,
the models yield clearly different results. The corresponding
growth time is shorter than the cycle period, which invalidates the
normal-form approach. Furthermore, the Babcock-Leighton model has
entered a strongly nonlinear and presumably chaotic regime. While this
indicates the limits of that model, the similarity in the other cases
demonstrates its validity for not too strong dynamo excitation. This is
the realistic case for the Sun, but we expect more rapidly rotating and
very active stars to be in a more strongly nonlinear or even chaotic
regime \citep[see, e.g.,][]{Tobias:etal:1995}.

\section{Conclusion}
\label{sec:conclusion}

Our results suggest that the variability of the solar cycle amplitudes
between decadal and millennial timescales can be understood in terms of a
weakly nonlinear and noisy limit cycle. This approach is motivated by
observational results and its parameters are constrained by observations
as well. It represents the generic model for the fundamental mode of a
weakly excited $\alpha\Omega$-dynamo, such as the observationally well
supported Babcock-Leighton approach. Owing to its simplicity, the model
does not cover the possible forcing of higher dynamo modes (such as the
quadrupole mode leading to hemispheric asymmetry) and also does not
account for the deviations of the solar cycle from a (rectified)
sinusoidal shape. On the other hand, our results show that the long-term
variability of solar activity is consistent with fluctuations due to a
stochastic process, such as random scatter in the tilt angles of bipolar
magnetic regions and sunspot groups. No intrinsic periodicities apart
from the 11-year cycle are required to understand the variability,
although the possible existence of such periodicities cannot be strictly
excluded by our analysis.

\begin{acknowledgements}
The data for the Group Sunspot Numbers were obtained from the SILSO data
base maintained by the Royal Observatory of Belgium, Brussels
(http://sidc.oma.be/silso/groupnumberv3).  I. Usoskin kindly provided
the sunspot number reconstruction from the cosmogenic isotope record
presented in \citet{Usoskin:etal:2016}.
\end{acknowledgements}

\bibliography{variability} 
\bibliographystyle{apj}

\clearpage
\newpage

\clearpage
\newpage

\clearpage

\clearpage
\newpage

\clearpage
\newpage

\clearpage
\newpage

\clearpage
\newpage

\clearpage
\newpage

\clearpage
\newpage

\end{document}